\documentclass{aa}
\usepackage{latexsym}
\usepackage{amsmath}
\usepackage{amssymb}
\usepackage{graphicx}
\usepackage[english]{babel}

\begin{document}

\title{A strong case for fast stellar rotation at very low metallicities}

\author{C. Chiappini\inst{1,2}, R. Hirschi\inst{3}, G. Meynet\inst{2},
  S. Ekstr\"om\inst{2} A. Maeder\inst{2} \and F.Matteucci\inst{4}}

\institute{Osservatorio Astronomico di Trieste, Via G. B. Tiepolo 11, I - 34131 Trieste, Italia
\and Observatoire Astronomique de l'Universit\'e de Gen\`eve, CH-1290, Sauverny, Switzerland
\and Dept. of Physics and Astronomy, University of Basel, CH-4056, Basel, Switzerland
\and Dipartimento di Astronomia, Universit\'a degli Studi di Trieste, 
Via G. B. Tiepolo 11, I - 34131 Trieste, Italia}

\date{Received / Accepted}

\abstract{We investigate the effect of new stellar models taking 
  rotation into account and computed
  for a metallicity Z = 10$^{-8}$ on the chemical evolution
  of the earliest phases of the Milky Way. These models were computed
  under the assumption that the ratio of the initial rotation velocity
  to the critical velocity of stars is
  roughly constant with metallicity. This naturally leads to faster 
rotation at lower metallicity, as metal-poor stars are more compact
  than metal-rich ones. We find that the new Z = 10$^{-8}$
  stellar yields have a tremendous impact on the
  nitrogen enrichment of the interstellar 
medium for log(O/H)+12 $<$ 7 (or [Fe/H]$<$ $-$3). 
We show that upon the inclusion of the Z = 10$^{-8}$ stellar
  yields in chemical evolution models, both high N/O and C/O ratios
  are obtained in the very-metal poor metallicity range, in agreement
  with observations. Our results
  give further support to the idea that stars at very low
  metallicities could have rotational velocities of the order of 600-800\,km\,s$^{-1}$.

}

   \authorrunning{C. Chiappini et al.}

   \titlerunning{The Faster rotating stars at low metallicities}

\maketitle

\section{The still unexplored very low metallicity range}

The extremely metal-poor halo stars ([Fe/H] $<$ $-$3,
log(O/H)+12$<$ 7 , Cayrel et al. 2004, Beers \& Christlieb 2005, Spite et al. 2005, hereafter S05) are not only
relics of the earliest phases of the formation of the Milky Way (MW), but
also represent the lowest metallicities measured in the
universe. So far, high-z observations were not able to find objects in
this very low metallicity range.
Therefore, it is in the halo of the MW that we
have the unique possibility of studying the evolution of abundance ratios
at metallicities never explored before.

In particular, until very recently the nitrogen enrichment in low
metallicity environments could be studied only through
HII region abundances in the outer parts of spiral galaxies or in 
blue compact galaxies (both showing log(O/H)+12 $\gtrsim$ 7, e.g.
Izotov et al. 2005) or
Damped Lyman Alpha systems (DLAs, [Fe/H] $>-$2.5, Wolfe et al. 2005). 
The very metal-poor halo stars of S05
are unique since they provide the opportunity to study the nitrogen enrichment of
the insterstellar medium (ISM) at metallicities never explored before in chemical evolution models\footnote{In this Letter we concentrate on the so-called {\it normal} very metal-poor
stars, since our goal is to explain the mean ISM enrichment at this
low metallicity range. Thus, we will not address the
carbon-rich-ultra-metal-poor stars - see Beers \& Christlieb 2005 for a
recent review on this subject.}.
New effects are observed at such low metallicities. Recent
measurements of nitrogen abundances in metal-poor stars (S05)
show a high N/O ratio suggesting high levels of
production of primary nitrogen in massive stars. 
Moreover, the N/O abundance ratios in metal-poor
stars exhibit a large scatter (roughly 1 dex, much larger than
their quoted error bars), although none of the stars measured so 
far has N/O ratios as low as the ones observed in DLAs.

In a recent paper Chiappini et al. (2005, hereafter CMB05) studied the
implications of this new data set on our
understanding of nitrogen enrichment in the MW.
By the time the latter paper was published there was no
set of stellar yields able to explain 
the very metal-poor data of S05. 
Using the so-called population III stellar yields available 
in the literature did not solve the problem either (see also Ballero et
al. 2006 for chemical evolution models computed with Pop. III stellar
yields provided by different authors). 
In CMB05 it was concluded that the only way to account for
the new data was to assume that stars at low metallicity rotate sufficiently
fast to enable massive stars to contribute much larger 
amounts of nitrogen.
It was predicted that massive stars born with
metallicities below Z=10$^{-5}$ should produce a factor of 10 up to
a few times 10$^2$ more nitrogen (depending on the stellar mass) than the values 
given by Meynet \& Maeder (2002a, hereafter MM02) for Z=10$^{-5}$ and
$\upsilon^{\rm ini}_{\rm rot}=300$\,km\,s$^{-1}$.
The physical motivation for this 
would be an increase in the rotational velocity in very metal-poor
stars (Maeder et al. 1999; Meynet et al. 2005) and hence an increase in the nitrogen
yields (MM02 - see below). 
In this framework it is possible to understand the apparently
 contradictory
finding by S05 of a large scatter in N/O and the
 almost complete lack of scatter in [$\alpha$/Fe] ratios of the same
 very metal-poor halo stars (Cayrel et al. 2004).
In fact, the scatter in the N/O abundance ratios could
be related to the distribution of the stellar rotational
velocities as a function of metallicity.

Whether the above suggestions were physically plausible remained to 
be assessed by stellar evolution models computed at lower
metallicities, which took rotation and mass loss into account.
New stellar evolution models have been 
computed for metallicities Z=10$^{-8}$ (Meynet et al. 2005, Hirschi
2006) for massive stars. 
The new calculations show that, if the stars at all Z start their
lives on the zero age main sequence (ZAMS) with on average a
fraction of $\sim$\,0.5 the
critical velocity, the low Z stars easily reach break-up velocity
during MS evolution. Fast rotation also contributes producing a more efficient
mixing at lower Z, thereby leading to a large production of N in massive
stars. 
In this Letter we show the impact of
these new stellar yields on the chemical enrichment of C, N, and O of the earliest
phases of the MW. The new stellar yields are presented in
Sect. 2. Our results are shown in Sect. 3 and our conclusions drawn
in Sect. 4.

\section{New yields for very metal-poor massive stars}

This Letter focuses in the very metal-poor range (below
[Fe/H] $\simeq -$3 or log(O/H)+12 $\simeq$ 7) where no intermediate mass star
would have had time to contribute to the ISM enrichment (according to
our
non-stochastic approach). The effect of rotation on the evolution of 
low and intermediate mass stars (LIMS) and their impact on the galactic
ISM at higher metallicities will be discussed elsewhere.
Here we
adopt the stellar yields for LIMS given in MM02 but aware of the
limitations
already discussed in Chiappini et al. (2003). 

\begin{figure}
\centering
\includegraphics[width=5cm,angle=0]{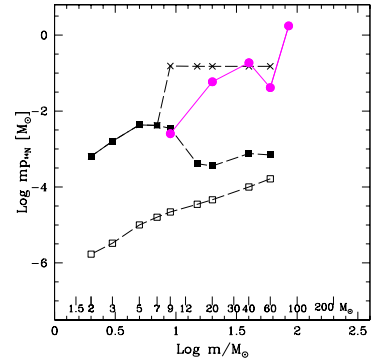}
\caption{Stellar yields for $^{14}$N at low metallicities for the
  whole stellar mass range. The yields of MM02 for stellar models with
  and without rotation for Z = 10$^{-5}$ are shown by filled and open
  squares, respectively. The asterisks connected by the long-dashed
  line show the {\it ad hoc} stellar yields adopted for Z$<$ 10$^{-5}$ in the
  heuristic model of CMB05 (see text). The dots show the new stellar
  yields computed for massive stars born at Z=10$^{-8}$. 
The new computations of Hirschi (2006) for Z=10$^{-8}$ lead to 
a large increase in N yields for masses above $\sim$20 M$_{\odot}$,
  similar to the predictions of CMB05. This agreement is striking
  considering that they were obtained from completely different approaches.}
\end{figure}

\begin{figure}
\centering
\includegraphics[width=5cm,angle=0]{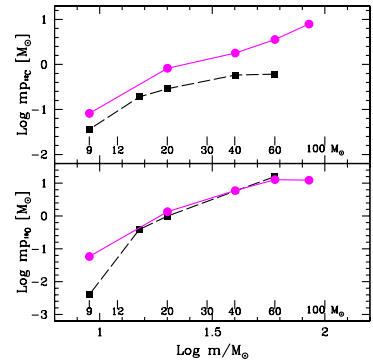}
\caption{Stellar yields for $^{12}$C (upper panel) and for  $^{16}$O (lower panel)
for massive stars. Curves labeled as in Fig. 1.}
\end{figure}

For massive stars we adopt the
new yields shown in Fig. 1. Filled
squares show stellar yields  of MM02 for their lowest
metallicity case (Z=10$^{-5}$) resulting from models with rotation
($\upsilon^{\rm ini}_{\rm rot}=300$\,km\,s$^{-1}$),
while open symbols stand for models computed with $\upsilon^{\rm ini}_{\rm rot}=0$\,km\,s$^{-1}$. The asterisks connected by the long-dashed line show
the {\it ad hoc} stellar yields for metallicities Z$<$10$^{-5}$
adopted in the {\it heuristic model} of CMB05. The dots show the
new results obtained by Hirschi (2006; see also Meynet et al. 2005)
for Z=10$^{-8}$.
The stellar yields for the Z=10$^{-8}$
case were computed according to the following assumption: stars begin
    their evolution on the ZAMS with approximately the same angular
momentum content, regardless of their metallicity. At solar metallicity,
    observations indicate a mean rotational velocity of a 60
    M$_{\odot}$ star on the MS of the order of 200\,km\,s$^{-1}$, which corresponds to an initial angular
momentum of the order of 2 $\times$ 10$^{53}$\,g\,cm$^2$\,s$^{-1}$ (see
    Meynet et al. 2005 for details). 
At a metallicity of Z=10$^{-8}$, this corresponds to
$\upsilon^{\rm ini}_{\rm rot}=$ 800\,km\,s$^{-1}$. In other words, we adopt a rotational
velocity such that
    the ratio between $\upsilon^{\rm ini}_{\rm rot}/\upsilon_{\rm breakup}$ remains almost constant
    (around 0.5) with mass and metallicity (see Hirschi 2006 for details).
The very interesting result is that the new computations by Hirschi (2006) for Z=10$^{-8}$ predict a large increase in the N
    yields for stars above 20 M$_{\odot}$, similar to the {\it ad hoc}
    yields
of CMB05 (see Fig.1).
Figure 2 shows a comparison of the stellar yields from stellar evolution
models with rotation for Z=10$^{-8}$ ($\upsilon^{\rm ini}_{\rm rot}=$
800\,km\,s$^{-1}$) and 10$^{-5}$ ($\upsilon^{\rm ini}_{\rm rot}=$ 300\,km\,s$^{-1}$) for $^{12}$C
(upper panel) and $^{16}$O (lower panel). The carbon yields for the
Z=10$^{-8}$ case are systematically higher than the ones at Z=10$^{-5}$ (although by significantly
lower factors than in the case of $^{14}$N - see Sect. 3).
The differences between the Z=10$^{-8}$ and 10$^{-5}$
yields shown in Figs. 1 and 2 are to be ascribed mainly to the
different $\upsilon^{\rm ini}_{\rm rot}$ adopted in each case, the physics
involved being essentially the same.

\begin{figure}
\centering
\includegraphics[width=6cm,angle=0]{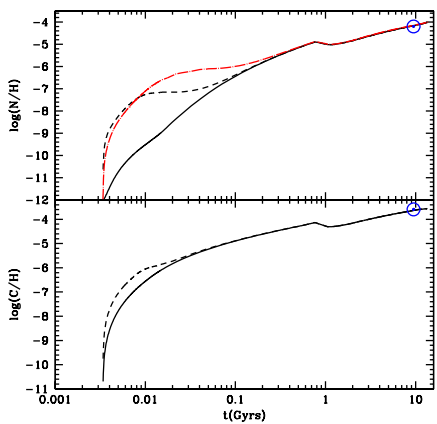}
\caption{Predicted evolution of nitrogen (upper panel) and carbon
  (lower panel) according to a chemical evolution model computed with
  different stellar yield sets for metallicities below Z=10$^{-5}$ : 
  a) solid line - a model computed under
  the assumption that the lowest metallicity yield table of MM02
  (Z=10$^{-5}$) is valid
  down to Z=0; b) dot-dashed line (red in the
  online version) - the model of 
CMB05 where an {\it ad hoc} higher yield of nitrogen is assumed for
  metallicities below 10$^{-5}$; c) dashed line - the new
  model presented in this Letter adopting new stellar yields
  computed by new stellar evolution models with faster rotation for
  Z=10$^{-8}$ massive stars.}
\end{figure}

\begin{figure}
\centering
\includegraphics[width=8cm,angle=0]{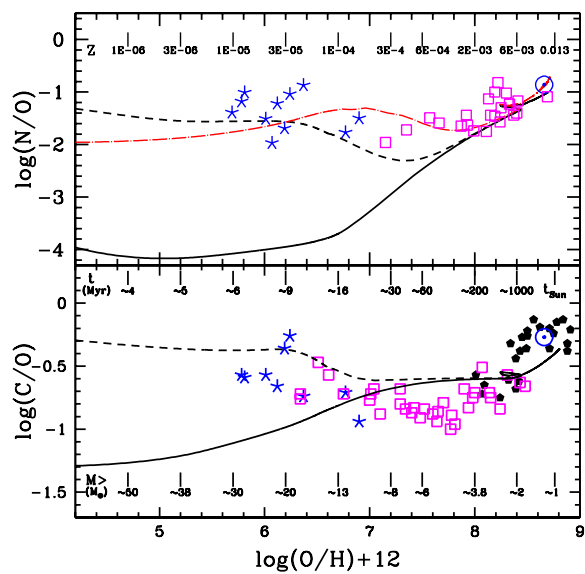}
\caption{Upper panel: Solar vicinity diagram log(N/O)
  vs. log(O/H)+12. The data points are from Israelian et al. (2004 - large
  squares), S05 (asterisks). Models are labeled as in Fig. 3.
Lower panel: Solar vicinity diagram log(C/O)
vs. log(O/H)+12. The data are from Spite et al. (S05 - asterisks),
Israelian et al. (2004 - squares), Nissen (2004 - filled pentagons). Solar
abundances (Asplund 2005 and references
therein) are also shown.}
\end{figure}

\section{Chemical evolution at very low metallicities}

The adopted chemical evolution model for the MW is the so-called
``two-infall model'', which is the same model as was used in CMB05.
Figure 3 shows our model predictions for the evolution of
nitrogen and carbon for different assumptions of stellar yields at low
metallicity.
The solid curves show a model computed with MM02 stellar yields
(as in Chiappini et al. 2003
\footnote{For the solar metallicity we are adopting the new
    calculations presented in Hirschi et al. (2004).}).
This model was computed under the
assumption that the Z=10$^{-5}$ yield table ($\upsilon^{\rm ini}_{\rm rot}=300$\,km\,s$^{-1}$) would be valid down to
Z=0. As can be seen in Fig. 4, this model cannot
explain the high levels of N/O or the C/O upturn observed in
the very metal-poor halo stars of S05\footnote{One of the main assumptions when comparing chemical evolution predictions with abundance
data is that they represent the pristine abundances from the ISM, from
which the stars formed. Therefore, objects that could have undergone
mixing processes should be avoided. The data shown here are in
principle {\it unmixed stars} (S05).}.
The dot-dashed line shows the {\it
  heuristic model} of CMB05. 
This model is the same as the one represented by the 
solid line except that for metallicities Z $<$ 10$^{-5}$ the 
yields of nitrogen were strongly increased in comparison to the
ones given in MM02 for massive stars (the adopted yields
in the case of this model are shown in Fig. 1 by the asterisks
connected by a long-dashed line).
As a consequence, this model produces more nitrogen
at the beginning of galaxy evolution (as shown in Fig. 3, upper panel,
dot-dashed curve), 
leading to large N/O ratios at low metallicities (Fig. 4, upper panel,
dot-dashed curve). However, it was unclear whether stellar
evolution models at such low metallicities could predict such a large
enhancement of nitrogen and what would be the impact for C and O.

The dashed curves in Figs. 3 and 4 show our most recent model computed with the new stellar
yields at Z=10$^{-8}$ of Hirschi (2006) for massive stars,
assuming them to be valid down to Z=0\footnote{The physics adopted in 
the Z$=$10$^{-8}$ models should be valid down to Z$\sim$10$^{-10}$,
which represents the metallicity limit below which massive stars first
enter the phase of H-burning via the pp chain, followed by the 3
$\alpha$ reaction, which then allows the CNO cycle to proceed, 
as in Z=0 (Pop. III) stars.}. 
The evolution of nitrogen predicted by this model (dashed
line in Fig. 3, upper panel) is similar to the one predicted by 
the CMB05 model (dot-dashed line), except
for the later times (as expected since the {\it ad hoc} yields of CMB05
are higher than the ones of Hirschi for masses below 20
M$_{\odot}$). The similarity of both curves is striking as
they were obtained following completely independent approaches.
This result implies that faster rotation is able to account for the
S05 observations. 
Some differences are also seen
for the C evolution (see Fig. 3, lower panel).
Our new model predicts a C/O upturn at low
metallicities (Fig. 4, lower panel, dashed curve).
This upturn results from the strong production of primary nitrogen.
Indeed, high production of primary nitrogen implies a very active H-burning shell, which
contributes a large part of the total luminosity of the star. As a consequence, part of
the total luminosity compensated by the energy produced in the helium core is reduced, making the average core temperature
and thus the efficiency of the $^{12}$C($\alpha$,$\gamma$)$^{16}$O
reaction lower. More efficient mixing also leads to greater mass loss,
decreasing the He-core size. Higher C/O ratios are thus obtained at the end of the He-burning phase.

We note that the effects obtained in this
Letter, namely, the high N/O and the C/O upturn at very low
metallicities can be explained without invoking Pop. III stars (i.e
without changing the IMF or including zero-metallicity stellar
yields) and hence do not necessarily imply the signature of Pop. III stars as
previously claimed in the literature (e.g. Akerman et al. 2004).
The only shortcoming of the present model is that it predicts
slightly higher [C/Fe] with respect to the observations of S05 (around
0.3 dex larger at [Fe/H] $\sim -$3 up to 0.9 dex at [Fe/H] $\sim -$4). 
However, in this case there are two large uncertainties, namely
{\it a)} the stellar yields for Fe in massive stars (strongly
dependent
on the adopted mass cut) and {\it b)} the [C/Fe] ratios of S05 could
still be affected by uncertainties due to the first dredge-up
dilution of C on the giant branch (Iben 1965) and NLTE/3D corrections.

Intermediate mass stars of this same low metallicity could also produce large amounts
of nitrogen, if this production is linked to high rotational
velocities. In this case, a flatter curve in the N/O
diagram would be obtained. In fact, as can
be seen in Fig. 4, the agreement between model predictions for N/O and C/O 
observations worsens at log(O/H)+12 $>$ 7, where the AGB contribution 
starts to be effective.
Super AGB stars, not included in the present
  Letter, could also contribute to the N enrichment of the ISM (Siess,
  priv. comm.).
Stellar evolution models with rotation for LIMS 
computed up to the final evolutionary
phases are being developed and will soon be available to be
incorporated
in chemical evolution models.

\section{Discussion and conclusions}

In this Letter, we computed chemical evolution models adopting the 
very recent calculations of Hirschi (2006)
for the evolution of massive stars at very low metallicities
under
the assumption of an almost constant ratio $\upsilon^{\rm ini}_{\rm rot}/\upsilon_{\rm breakup}$ as a
function of metallicity (i.e. where the $\upsilon^{\rm ini}_{\rm rot}$ increases towards lower
metallicities). In such a framework,
massive stars can produce large amounts of nitrogen (of the order of
the ones suggested in CMB05). 

At present, this is the only way to
explain the high nitrogen abundances
 measured recently in {\it normal} halo stars (S05).
This gives further support to the idea that 
stars rotate faster at very low metallicities.
The new stellar evolution models
also produce some extra carbon at Z $\leq$ 10$^{-5}$. As a
consequence, an upturn is produced in the C/O at log(O/H)+12 $<$ 7
metallicities. This result is obtained without the need of introducing
Pop. III stars. Finally, our results show that
the C and N yields strongly depend on metallicity for Z $\leq$ 10$^{-5}$, once rotation is
 taken into account. This has a strong impact on the chemical enrichment
 of the very metal-poor ISM. It is thus important to avoid extrapolating the stellar
 yields in this metallicity range when computing chemical evolution
 models, even if the timespan to reach Z=10$^{-5}$ is just a few
 Myrs.

The only alternative explanation of the observations
 considered in this Letter would be if AGB stars contributed to
 the ISM enrichment before [Fe/H] $\sim$ $-$3. This would mean that 
 the timescales for the
 chemical evolution of the halo would be very different from the
 ones adopted here. Our model assumes that the halo was formed by
 an exponential infall with an e-folding time of $\sim$1Gyr. The
 enrichment timescales do not change if the infall timescale is
 increased
by a factor of 2 (nor if the star formation efficiency is divided by 2).
Models assuming outflows could delay the ISM enrichment so that AGBs
 would have time to contribute. However, in this case
 it should be
seen if such models could still fit the halo abundance ratios well.  
The chemical evolution model presented here (without outflows) 
also agrees well with several
 other abundance ratios (Fran\c cois et al. 2004). 

Here we show the results for the mean ISM enrichment history, assuming that at a given
metallicity all stars of a given mass rotate at the same
velocity. Further computations are in progress, and we envision
being able to compute chemical evolution models in which a
distribution
of rotational velocities is assumed for each metallicity. These models
would predict  
a scatter in N/O to be compared to the observed one (forthcoming paper).


\end{document}